\documentclass[aps,prl,twocolumn,floatfix,a4paper,showpacs,superscriptaddress,eqsecnum]{revtex4}
\usepackage{epsfig}
\usepackage{amsmath}
\usepackage{color}

\newcommand{\psinit}{\psi_\mathrm{init}}
\newcommand{\ii}{\mathrm{i}}
\newcommand{\e}{\mathrm{e}}
\newcommand{\op}[1]{\hat{#1}}
\newcommand{\order}{{O}} 
\newcommand{\opt}{m_{\mathrm{opt}}} 
\newcommand{\ke}[1]{\left|{#1}\right>}

\newcommand{\Ctrl}{\mathcal C\op V_f}
\newcommand{\CtrlU}{\mathcal C\op U_f}
\newcommand{\CtrlWone}{\mathcal C\op W_1}
\newcommand{\CtrlWtwo}{\mathcal C\op W_2}

\newcommand{\subspace}{{\mathcal S}}

\newcommand{\comment}[1]{{\color{red}#1}}
\renewcommand{\comment}[1]{}

\begin{document}
\title{Quantum searches on highly symmetric graphs}

\author{Daniel Reitzner}
\affiliation{Research Center for Quantum Information, Slovak Academy of Sciences, D\'ubravsk\'a cesta 9, 845 11 Bratislava, Slovakia}
\affiliation{Quniverse, L\'\i\v s\v cie \'udolie 116, 841 04, Bratislava, Slovakia}
\author{Mark Hillery}
\affiliation{Department of Physics, Hunter College of the City University of New York, 695 Park Avenue, New York, NY 10021}
\author{Edgar Feldman}
\affiliation{Department of Mathematics, Graduate Center of the City University of New York, 365 Fifth Avenue, New York, NY 10016}
\author{Vladim\'\i r Bu\v zek}
\affiliation{Research Center for Quantum Information, Slovak Academy of Sciences, D\'ubravsk\'a cesta 9, 845 11 Bratislava, Slovakia}
\affiliation{Quniverse, L\'\i\v s\v cie \'udolie 116, 841 04, Bratislava, Slovakia}

\begin{abstract}
We study scattering quantum walks on highly symmetric graphs and use the walks to solve search problems on these graphs. The particle making the walk resides on the edges of the graph, and at each time step scatters at the vertices. All of the vertices have the same scattering properties except for a subset of special vertices. The object of the search is to find a special vertex. A quantum circuit implementation of these walks is presented in which the set of special vertices is specified by a quantum oracle. We consider the complete graph, a complete bipartite graph, and an $M$-partite graph. In all cases, the dimension of the Hilbert space in which the time evolution of the walk takes place is small (between three and six), so the walks can be completely analyzed analytically. Such dimensional reduction is due to the fact that these graphs have large automorphism groups. We find the usual quadratic quantum speedups in all cases considered.
\end{abstract}

\maketitle

\section{Introduction}
\label{sec:intro}

The theory of quantum walks describes quantum versions of classical random walks. In these walks a quantum particle ``walking'' on a line, or more generally a graph, has different amplitudes to go in different directions rather than different probabilities, which is the case for a classical walk. The time in these walks can either be in discrete steps \cite{AhDaZa93,AhAmKeVa01} or continuous \cite{FaGu98b}. Both types of walks have proven to be fruitful sources of quantum algorithms \cite{ambainis07,ChEi05,ChClDeFaGuSp03,FaGoGu07}. A summary of both the properties of quantum walks and their algorithmic applications can be found in two recent reviews \cite{kempe03,kendon07}.

Quantum walks have been used to investigate searches on a number of different graphs. In these searches, one of the vertices is distinguished (special), and the object is to find that vertex. The graphs considered so far are grids and hypercubes of different dimensions and the complete graph \cite{ShKeWh03,AmKeRi05,ChGo04}.

In this paper, we further pursue the possibility of using quantum walks to speed up searches on graphs. We shall consider highly symmetric graphs, that is, those having a large automorphism group. As we shall see, on these graphs the quantum walk takes place in a subspace of small dimension, and this greatly simplifies the analysis of its behavior. The role of symmetry in quantum walks was studied explicitly in Ref.~\cite{KrBr07}. The authors of this paper considered coined quantum walks on Cayley graphs of groups and found that in some cases a quantum walk on a large graph could be reduced to one on a much simpler graph, which they called a quotient graph. The situation for a search on a highly symmetric graph is analogous to that in the Grover search algorithm \cite{grover97} in which the search takes place in a two-dimensional subspace. The result is that it is quite simple to find the behavior of the quantum search and compare it to the classical one.

We shall consider searches on complete graphs and bipartite graphs, and an $M$-partite graph, that includes both the complete graph and the bipartite graphs as special cases. We shall make use of the scattering quantum walk \cite{HiBeFe03}. In addition, we shall show how a quantum search on a graph can be implemented by a quantum circuit containing a quantum oracle. The oracle is used to differentiate the properties of the special vertex from those of the normal vertices.

The rest of the paper is organized as follows. We start in Sec.~\ref{sec:SQW} by introducing scattering quantum walks and show that they can be realized with a quantum circuit. A quantum
walk search can be implemented by adding a quantum oracle to the circuit (Sec.~\ref{sec:connection}). In Sec.~\ref{sec:symmetry} we show how symmetry of the underlying graph reduces the dimensionality of the problem, and then, in Sec.~\ref{sec:examples}, we apply these observations to different graphs: a complete graph, a bipartite graph, and a special $M$-partite complete graph. We also show how the behavior of the search on the complete graph depends on the phase shift that the special vertex imparts to the particle upon scattering it. At the end of the section we also present a comparison of our results with other works. We summarize our results in Sec.~\ref{sec:conclusion}).

\section{Scattering quantum walks}
\label{sec:SQW}

The scattering quantum walk, introduced in Ref.~\cite{HiBeFe03}, is a quantum walk on the edges of a graph rather than on its vertices. Having a graph $\mathcal G=(V,E)$ on which the walk is defined, with $V$ being the set of vertices and $E$ the set of edges, the Hilbert space is defined as
\begin{equation}
\label{eq:hilbik}
\mathcal H=\ell^2(\{\ke{m,l}|m,l\in V,ml\in E\}).
\end{equation}
This definition states that the Hilbert space is given by the span of all \emph{edge states}, i.e., position states $\ke{m,l}$ interpreted as a particle going from vertex $m\in V$ to vertex $l\in V$, with $ml\in E$ being an edge of graph $\mathcal G$. These edge states form an orthonormal basis of the Hilbert space, which we shall call \emph{the canonical basis.}

In this Hilbert space the unitary evolution is given by a set of \emph{local unitary evolutions} defined for each vertex. If we set (for every $m\in V$) $A_m=\ell^2(\{\ke{m,l}|l\in V, ml\in E\})$, the set of all edge states originating on vertex $m$, and $\Omega_m=\ell^2(\{\ke{l,m}|l\in V, lm\in E\})$, the set of all edge states ending on vertex $m$, then local unitary evolutions act as $\op U^{(m)}:\Omega_m\to A_m$. The overall unitary step operator $\op U$ acting on the system is  the combined action of the local unitary evolutions, that is, the restriction of $\op U$ to $\Omega_{m}$ is just $U^{(m)}$. Given the initial state of the system is $\ke{\psinit}$, the state after $n$ steps is $\ke{\psi_n}=\op U^n\ke\psinit$ and the probability of finding the particle (walker) in state $\ke{k,l}$ is then $|\langle k,l|\psi_n\rangle|^2$.

Finally, let us specify our choice of local unitary step operators.  Our vertices will be of two types, normal and special, and our task will be to find at least one special vertex. Consider a vertex $l$, let $\Gamma (l)$ be the set of vertices connected to $l$ by an edge, and if $k\in \Gamma (l)$, let $\Gamma (l;k)$ be the set of vertices connected to $l$ by an edge but excluding $k$.  The local unitary operators corresponding to both the normal and special vertices will act as follows:
\begin{equation}
\op U^{(l)}\ke{k,l}=-r^{(l)}\ke{l,k}+t^{(l)}\sum_{m\in \Gamma (l;k)}\ke{l,m},
\label{eq:unitary}
\end{equation}
where $r^{(l)}$ and $t^{(l)}$ are reflection and transmission coefficients to be chosen in such a way that $\op U^{(l)}$ is unitary. For normal vertices
\begin{equation}
t^{(l)}=\frac{2}{|\Gamma(l)|},\qquad r^{(l)}=1-t^{(l)}, 
\label{eq:GC}
\end{equation}
and for special vertices, we choose
\begin{equation}
\label{eq:RC}
r^{(l)}=-\e^{\ii \phi},\qquad t^{(l)}=0 .
\end{equation}
Here $|\Gamma(l)|$ is the degree of vertex $l$, i.e., the number of vertices in the set $\Gamma (l)$.  For both of these choices, the operator $\op U^{(l)}$ is unitary, and, as we shall see, they also guarantee that the quantum walk has the same symmetry group as the graph.  This choice of local unitary operators for the scattering walk is analogous to the choice of the Grover coin (see Ref.~\cite{MaBaStSa02}) in a coined quantum walk.

\section{Quantum circuit settings}
\label{sec:connection}

Quantum search problems are often posed in terms of a quantum oracle.  This is a black box that, when given an input value $x$, returns an output $f(x)$.  The function takes values $0$ and $1$, and the set of inputs for which it takes the value $1$ is typically small.  The objective is to find at least one input for which the function is equal to $1$.  What we would like to do in this section, is provide a connection between this phrasing of a search problem and a search whose object is to find a set of special vertices on a graph.  We shall accomplish this by showing how an oracle can be used in a quantum circuit to implement a quantum search on a graph.

Our oracle has the action
\begin{equation}
\label{eq:blackbox}
\Ctrl\ke k\otimes\ke m=\ke k\otimes\ke{m\oplus f(k)},
\end{equation}
where $k$ is the input and $f$ is the function that determines the action of the oracle. We shall assume that $f(k)$ is either $0$ or $1$, though this restriction is not necessary.  The second system (ancilla) is, therefore, a qubit, and $\oplus$ is addition modulo $2$. In addition to this operator, we need one that will link  the values of the function to the action of local unitary evolutions at the vertices. In particular, if $l$ is a vertex of the graph, the $f(l)=0$ will correspond to a normal vertex and $f(l)=1$ will correspond to a special vertex, i.e., the type of vertex we are trying to find.  If $l$ is a normal vertex, the local unitary operator corresponding to it will be denoted by $\op U_{0}^{(l)}$, and if it is a special vertex the local unitary will be denoted by $\op U_{1}^{(l)}$. Our quantum circuit will act on a tensor product of the Hilbert space $\mathcal H$ for the quantum walk [see Eq.~(\ref{eq:hilbik})], the Hilbert space for vertices, $\mathcal{H}_{v}$, and a qubit Hilbert space $\mathcal{H}_{2}$. The vertex space is given by
\begin{equation}
\mathcal{H}_{v} = \ell^2(\{\ke{l}| l\in V\}).
\end{equation}
We now define an operator, $\CtrlU$, acting on this space in the following way
\begin{equation}
\label{eq:walk}
\CtrlU(\ke{k,l}\otimes\ke l\otimes\ke c)=\left(\op U^{(l)}_{c}\ke{k,l}\right)\otimes\ke l\otimes\ke c.
\end{equation}
This equation does not completely specify the actions of $\CtrlU$. In particular, it does not specify its action on states of the form $|k,l\rangle \otimes |l^{\prime}\rangle \otimes |c\rangle$, where $l\neq l^{\prime}$, but we will only need to consider its action on states of the form given in the previous equation.

\begin{figure}
\includegraphics{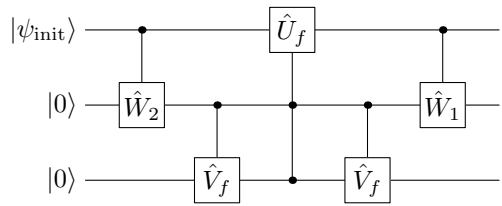}
\caption{\label{fig:circuit}
A logical circuit (network) that implements a single step of a scattering quantum walk search, which makes use of the quantum oracle $\Ctrl$. The first input corresponds to a quantum walker originally prepared in the state $\ke\psinit$. The second input represents a vertex
state, while the third input represent an ancillary qubit.}
\end{figure}

The quantum circuit that implements one step of our quantum walk search is given in Fig.~\ref{fig:circuit}. The first input stands for a state of quantum walker (i.e.~any superposition of edge states), the second for a vertex state, and the third for an ancillary qubit.  The input state is $\ke{\psi_0}=\ke\psinit\otimes\ke 0\otimes\ke 0$ where $\ke\psinit$ is a general state in the edge Hilbert space $\mathcal{H}$,
\[
\ke\psinit=\sum_{ml\in E}a_{ml}\ke{m,l},\qquad\sum_{ml\in E}|a_{ml}|^2=1.
\]
The state $|0\rangle$ in the second input in Fig.~\ref{fig:circuit} is one of the vertex states, which, besides labeling a particular vertex, will also serve as a reference state.  First, we apply the operator $\CtrlWtwo$ which maps the state $|m,l\rangle \otimes |0\rangle$ in $\mathcal{H} \otimes \mathcal{H}_{v}$ to $|m,l\rangle \otimes |l\rangle$.  Such a unitary can be implemented, e.g.~as presented in Refs.~\cite{BoBrHoTa96,BrBuHi01} and references therein.  After this  operator is applied, the initial state becomes
\[
\ke{\psi_0}\mapsto\sum_{ml\in E}a_{ml}\ke{m,l}\otimes\ke{l}\otimes\ke 0\equiv\ke{\psi_1}.
\]
Next, we apply the quantum oracle given by Eq.~(\ref{eq:blackbox}) to the vertex state and the qubit, yielding
\[
\ke{\psi_1}\mapsto\sum_{ml\in E}a_{ml}\ke{m,l}\otimes\ke{l}\otimes\ke{f(l)}\equiv\ke{\psi_2}.
\]
Now we can apply the $\CtrlU$ operator from Eq.~(\ref{eq:walk}) to the state, which gives us
\[
\ke{\psi_2}\mapsto\sum_{ml\in E}a_{ml}\left(\op U_{f(l)}^{(l)}\ke{m,l}\right)\otimes\ke{l}\otimes\ke{f(l)}\equiv\ke{\psi_3}.
\]
Here it is clear that both ancillary systems act as controls, and they will have to be reset before we can use the circuit again.  That is the task of the remaining two gates in the circuit.  Now, because the local unitary operator $\op U_{f(l)}^{(l)}$ acts only on the edge space, the state can be rewritten as
\[
\ke{\psi_3}=\sum_{lm\in E}b_{lm}\ke{l,m}\otimes\ke{l}\otimes\ke{f(l)}.
\]
Application of the quantum oracle again resets the qubit state to $|0\rangle$. However, to reset the vertex state, we cannot use $\CtrlWtwo$ operation as before, since the information contained in the edge state about the vertex state has moved from the second to the first slot.   Therefore, we define the operator $\CtrlWone$, which maps $|l,m\rangle \otimes |l\rangle$ in $\mathcal{H} \otimes \mathcal{H}_{v}$ to $|l,m\rangle \otimes |0\rangle$, and apply it to our state, giving
\[
\ke{\psi_3}\mapsto\sum_{lm\in E}b_{lm}\ke{l,m}\otimes\ke{0}\otimes\ke{0}.
\]
We have thus performed one step of the walk and reset the ancillas, so that the circuit can be applied again to perform additional steps of the walk.

\section{Symmetry considerations}
\label{sec:symmetry}

For simplicity, suppose we have a graph $\mathcal G=(V,E)$ with only one special vertex, and let $\mathcal{A}$ be the group of automorphisms of the graph that leave the special vertex fixed.  An automorphism $a$ of $\mathcal G$ is a mapping $a: V\rightarrow V$ such that for any vertices $v_{1},v_{2} \in V$ there is an edge connecting $a(v_{1})$ and $a(v_{2})$ if and only if there is an edge connecting $v_{1}$ and $v_{2}$.  Each automorphism, $a$, induces a unitary mapping $\op U_a$ on the Hilbert space of the graph $\mathcal G$ given by Eq.~(\ref{eq:hilbik}), such that $\op U_{a}|v_{1},v_{2}\rangle =|a(v_{1}),a(v_{2})\rangle$.  Suppose now that $\mathcal{H}$ can be decomposed into $m$ subspaces,
\[
\mathcal{H} = \bigoplus_{j=1}^{m} \mathcal H_{j} ,
\]
where each $\mathcal H_{j}$ is the span of some subset $B_{j}$ of the canonical basis elements and is invariant under $\op U_{a}$ for all $a \in \mathcal{A}$.  We shall also assume that each $\mathcal H_{j}$ does not contain any smaller invariant subspaces $m$.

Next, in each invariant subspace we can form a vector that is the sum of all of the canonical basis elements in the subspace,
\begin{equation}
\label{eq:eigenone}
|w_{j}\rangle = \frac{1}{\sqrt{d_{j}}} \sum_{|v_{1},v_{2}\rangle \in B_{j}}
 |v_{1},v_{2} \rangle  ,
\end{equation}
where $d_{j}$ is the dimension of $\mathcal H_{j}$.  This vector satisfies $\op U_{a}|w_{j}\rangle = |w_{j}\rangle$ for all $a \in \mathcal{A}$. Moreover, it is the only vector in $\mathcal H_{j}$ that satisfies this condition.  Define $\subspace= \ell^2( \{ |w_{j}\rangle | j=1,2, \ldots, m \})$, and note that $\subspace = \{ |\psi\rangle \in \mathcal{H} |\op U_{a}|\psi\rangle = |\psi\rangle\ \forall a\in \mathcal{A} \}$, and that the dimension of $\subspace$ is simply the number of invariant subspaces.  Now suppose that $[\op U,\op U_{a}]=0$ for all $a \in \mathcal{A}$.  This implies that if $\op U_{a}|\psi\rangle = |\psi\rangle$, then $\op U_{a}\op U|\psi\rangle = \op U|\psi\rangle$, and if $|\psi\rangle \in \subspace$, then $\op U|\psi\rangle \in \subspace$. Therefore, the subspace $\subspace$ is closed under the action of the step operator $\op U$. Correspondingly, if the initial state of the walk is in $\subspace$, then we only need to consider states in $\subspace$ to describe the state of the walk at any time.  If the automorphism group is large, then $\subspace$ can have a much smaller dimension than $\mathcal{H}$.

Now let us demonstrate that the unitary operator $\op U$, defined by the local unitary operators in Eq.~(\ref{eq:unitary}), does, in fact, commute with all of the automorphisms of a graph that leave the special vertex fixed.  If these operators commute when applied to all of the elements of the canonical basis, then they commute. As before, $\Gamma (v)$ is the set of vertices in $V$ that are connected to the vertex $v$, and, if $v^{\prime} \in \Gamma (v)$ then $\Gamma (v; v^{\prime}) = \Gamma (v) - \{v^{\prime}\}$ and, finally, $|\Gamma(v)|$ is the number of elements in $\Gamma(v)$. Then we have that
\begin{eqnarray*}
\op U_{a} \op U |v_{1},v_{2}\rangle & = & -r^{(v_2)} |a(v_{2}),a(v_{1})\rangle\nonumber \\
& & + t^{(v_2)}\!\!\! \sum\limits_{ v\in \Gamma (v_{2};v_{1})}\!\!\! |a(v_{2}),a(v)\rangle .
\end{eqnarray*}
We also have
\begin{eqnarray*}
\op U \op U_{a} |v_{1},v_{2}\rangle & = & -r^{[a(v_2)]} |a(v_{2}),a(v_{1})\rangle\nonumber\\
& & +  t^{[a(v_2)]} \sum\limits_{ v\in \Gamma (a(v_{2}); a(v_{1}))} |a(v_{2}),v \rangle .
\end{eqnarray*}
First note that the reflection and transmission amplitudes in this equation are the same as those in the previous equation, i.e., $r^{(v_2)}=r^{[a(v_2)]}$ and $t^{(v_2)}=t^{[a(v_2)]}$---this is a consequence of the fact that $| \Gamma (v_{2};v_{1})| = |  \Gamma (a(v_{2});a(v_{1})) |$ and the fact that the special vertex is mapped into itself by the automorphism.  We also have that $ \Gamma (a(v_{2});a(v_{1})) = \{ a(v)|\ v\in  \Gamma (v_{2};v_{1}) \}$, so that the sums in the two equations are identical.  Therefore, $U_{a} \op U |v_{1},v_{2}\rangle= \op U \op U_{a} |v_{1},v_{2}\rangle $.

Somewhat similar considerations appear in Ref. \cite{KrBr07}, but the situation there is complicated by the fact that the analysis was performed for a coined quantum walk.  This requires the edges to be colored, and the automorphisms to respect the coloring, that is,  edges must be mapped to edges of the same color.  The situation for the scattering quantum walk is much simpler, because no edge coloring is required.

\section{Examples of searches}
\label{sec:examples}

\subsection{Search on a complete graph}
\label{sec:cg}

\begin{figure}
\includegraphics{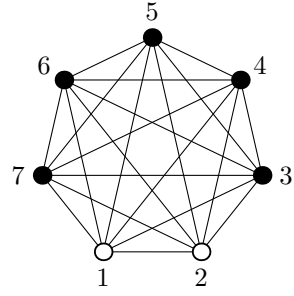}
\caption{\label{fig:CG}An example of a complete graph with $N=7$ vertices out of which $v=2$ are special (white ones). A solution for a scattering-quantum-walk search on such a graph leads to a reduction in dimensionality of the problem to four dimensions.}
\end{figure}

Let us consider a complete graph with $N$ vertices (see Fig.~\ref{fig:CG}). Each vertex of this graph is connected to all of the other vertices by an edge, so the graph has $N(N-1)/2$ edges which define the Hilbert space of dimension $\dim\mathcal H =N(N-1)$. Let $v$ be the number of special vertices. Then, without a loss of generality, we may label them as $k=1,2,\ldots, v$ and corresponding local unitary evolution is defined by Eqs.~(\ref{eq:unitary}) and (\ref{eq:RC}). Normal vertices are labeled as $k=v+1,v+2,\ldots, N$, and have local unitary evolution defined by Eqs.~(\ref{eq:unitary}) and (\ref{eq:GC}). Transmission and reflection coefficients for all normal vertices $a$ are the same, with $|\Gamma(a)|=N-1$.

The Hilbert space for the walk defined on this graph can be decomposed into four subspaces characterized by four vectors according to Eq.~(\ref{eq:eigenone}),
\begin{eqnarray*}
|w_{1}\rangle & = & \frac{1}{\sqrt{v(N-v)}}\sum_{a=v+1}^{N}\sum_{b=1}^v |a,b\rangle,  \nonumber \\
|w_{2}\rangle & = & \frac{1}{\sqrt{v(N-v)}}\sum_{a=1}^{v}\sum_{b=v+1}^N |a,b\rangle,  \nonumber \\
|w_{3}\rangle & = & \frac{1}{\sqrt{(N-v)(N-v-1)}}\sum_{a=v+1}^{N}\sum_{\substack{b=v+1\\b\neq a}}^{N} |a,b\rangle, \nonumber \\
\end{eqnarray*}
\begin{eqnarray*}
|w_{4}\rangle & = & \frac{1}{\sqrt{v(v-1)}}\sum_{a=1}^{v}\sum_{\substack{b=1\\b\neq a}}^{v} |a,b\rangle.
\end{eqnarray*}
These are equal superpositions of all edge states directed from normal vertices to special vertices, from special vertices to normal vertices, connecting normal vertices and connecting special vertices, respectively. Unitary evolution, as was shown in Sec.~\ref{sec:symmetry}, can be described by the evolution within the subspace $\subspace$ spanned by four vectors $\ke{w_k}$, $k=1,2,3,4$, and is described by a $4\times 4$ matrix
\begin{equation}
\label{eq:US}
\op U = \begin{pmatrix} 0 & q & s & 0 \\ \e^{\ii\phi} & 0 & 0 & 0 \\ 0 & s & -q & 0 \\ 0 & 0 & 0 & \e^{\ii\phi}
\end{pmatrix},
\end{equation}
where
\begin{eqnarray*}
q & = & -r+t(v-1)=-1+\frac{2v}{N-1},\nonumber\\
s & = & \sqrt{1-q^2}=t\sqrt{v(N-u-1)}.\nonumber
\end{eqnarray*}
Note that the subspace spanned by the vectors $|w_k\rangle, k=1,2,3$, is decoupled from the subspace spanned by the vector $|w_4\rangle$. For the initial state we choose an equal superposition of all edge states, i.e.,
\[
\ke\psinit=\frac{1}{\sqrt{N(N-1)}}\sum_{\ke{k,l}\in\mathcal H}\ke{k,l}.
\]
This initial state is a natural choice, since no set of edge states is preferred. We have that $\ke\psinit\in\subspace$ and can be written as a superposition of the states $\ke{w_k}$ for $k=1,2,3,4$,
\begin{multline*}
\ke\psinit = \sqrt{\frac{v(N-v)}{N(N-1)}}\left(\ke{w_1}+\ke{w_2}\right)\\
+ \sqrt{\frac{(N-v)(N-v-1)}{N(N-1)}}\ke{w_3}+\sqrt{\frac{v(v-1)}{N(N-1)}}\ke{w_4}.
\end{multline*}

We shall analyze what happens for two choices of a phase shift, namely, 
$\phi = 0$ and $\phi = \pi$.  For $\phi=0$, we find that the initial state can be written as a superposition of eigenstates of $\op U$ with eigenvalue equal to unity.  In particular, the following two states are eigenstates of $\op U$ with the eigenvalue equal to unity:
\begin{eqnarray*}
|\tilde u_0\rangle & = & \sqrt{\frac{1+q}{3+q}}\begin{pmatrix}
1 \\ 1 \\ \displaystyle{\sqrt{\frac{1-q}{1+q}}} \\ 0\end{pmatrix}, \nonumber \\
|\tilde u_0'\rangle & = & (0,0,0,1)^T ,
\end{eqnarray*}
and the initial state can be expressed in terms of them as
\begin{eqnarray*}
|\psinit\rangle & = & \sqrt{\frac{(N-v)(N+v-1)}{N(N-1)}}|\tilde u_0\rangle \nonumber \\
& & + \sqrt{\frac{v(v-1)}{N(N-1)}}|\tilde u'_0\rangle .
\end{eqnarray*}
This, however, means that the walk goes nowhere; the particle remains in the initial state for the entire walk. In this case, the quantum walk gives us no advantage over a classical search. Classical searches on edges are discussed in Appendix A.

The value of $\phi=\pi$ is much more interesting. The state of the walk after $n$ steps, derived by using the formula
\begin{equation}
\label{eq:evolution}
\ke{\psi_n}=\sum_{\lambda}\left(\lambda^n\langle u_\lambda|\psinit\rangle\right)\ke{u_\lambda},
\end{equation}
where $\ke{u_\lambda}$ is the eigenvector corresponding to eigenvalue $\lambda$ of unitary operator $\op U$, is, in the limit $N\gg v \geq 1$,
\begin{equation}
\label{eq:result}
\op U^n|\psinit\rangle\simeq\frac{1}{2\sqrt{Nv}}\begin{pmatrix}
\sqrt{2v(N-1)}\sin (2n+1)\displaystyle{\frac{\theta}{2}}\\
-\sqrt{2v(N-1)}\sin (2n-1)\displaystyle{\frac{\theta}{2}}\\
2\sqrt{v(N-v-1)}\cos n\theta\\
0
\end{pmatrix}, 
\end{equation}
where
\begin{equation}
\tan\theta = \frac{\sqrt{v(2N-v-2)}}{N-v-1}.
\end{equation}
We see that the probability amplitudes for edge states not connected to special vertices are approximately equal to zero when $\theta n=\pi/2$. If we measure the position of the particle after this many steps, it will, with probability close to unity, be located on an edge connected to one of the special vertices. This implies that the number of steps needed to find one of the special vertices is of the order $\order (\sqrt{N/v})$ for large $N$ which is a quadratic speedup over the classical algorithm that needs $\order (N/v)$ steps to do the task when searching an unstructured database.

It should be noted that for the case $v=1$ there is no vector $\ke{w_4}$ and the dimension of the problem is reduced to three, yet the results remain.

\subsubsection{Role of the phase shift}
\label{sec:optimality}

\begin{figure}
\includegraphics[scale=0.8]{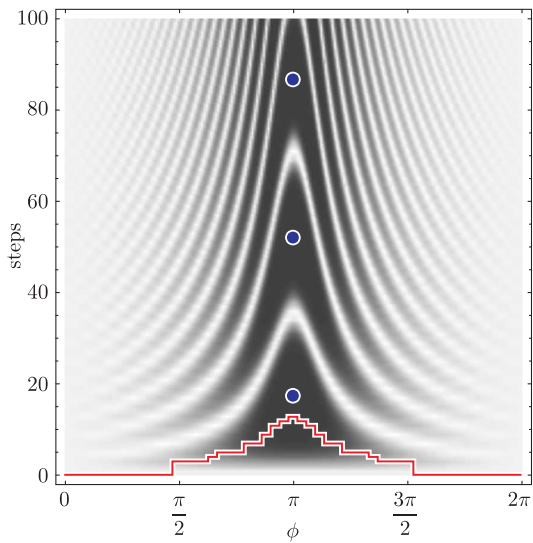}
\caption{\label{fig:density}(Color online) The density plot of the probability $P$, taken for a quantum walk on complete graph with $N=256$ and $v=1$, shows the convergence of the side ridges to the maximum (dark gray) for $\phi=\pi$ with the probability $P=1$ (blue dots). White areas correspond to the minimal probability of finding the particle on an edge connected to the special vertex. The thick red line represents the optimal number of steps that are to be taken before the measurement so that the search algorithm for the special vertex would require minimal average number of steps.}
\end{figure}

As we have seen in the previous section, the differences in the behaviors of quantum walks with a phase shift  $0$ and a phase shift $\pi$ are substantial. To see how this change occurs we will present numerical results for $0 \leq \phi < 2\pi$ for the case of a single special vertex. The behavior of the walk is depicted in Fig.~\ref{fig:density}, where we see, that the further we are from $\phi=\pi$, the smaller is the probability $P$ of finding the particle on an edge that is connected to the special vertex.

We can see from Fig.~\ref{fig:density} that $P$ is symmetric about $\phi =\pi$. This property can be shown in the following way. If we make the substitution $\phi\to2\pi-\phi$ in the unitary $\op U$ from Eq.~(\ref{eq:US}) we find that $\op U$ changes to $\op U^\ast$. This result together with the fact, that the initial state has real-valued coefficients when expanded in the canonical basis, yields
\[
\left(\op U^\ast\right)^n|\psinit\rangle=\left(\op U^\ast\right)^n|\psinit^\ast\rangle=\left(\op U^n|\psinit\rangle\right)^\ast.
\]
So the resulting components are now the complex conjugates of the original; however, the probabilities remain the same.

The value $\phi=\pi$ is special, because for any other $\phi$, the probability $P$ never reaches unity. The significance of $\phi=\pi$ leads to the following problem: Suppose we are given a complete graph having one special vertex with an arbitrary but known phase shift. Our task is to find the special vertex. In order to find it efficiently, we need to wait for the optimal number of steps before performing the measurement. If we were given only one chance to measure, we would have to wait until the probability of finding a particle on one of the edges connected to the special vertex reaches its maximum. However, if we are able to repeat the experiment an arbitrary number of times, we may, after an unsuccessful search, do the experiment again. In that case, the optimal number of steps before measuring is different---it is the number of steps, for which the \emph{average number of steps} is minimal.

\begin{figure}
\includegraphics[scale=0.9]{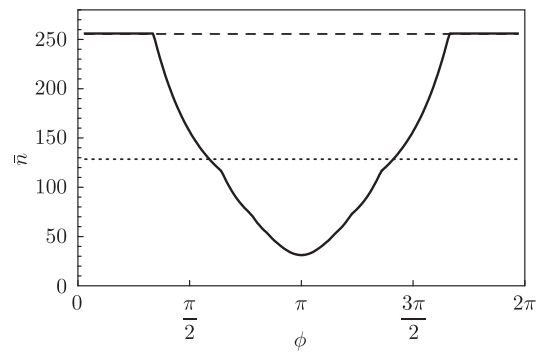}
\caption{\label{fig:optimality} Average number of walking steps as a function of the phase shift $\phi$. This plot illustrates the efficiency of algorithms---the classical blind search (dashed), the classical search with memory (dotted), and the quantum search (thick solid). We consider a graph with  $N=256$ vertices. We see that the minimal value of the average number of steps is achieved for the
quantum search with $\phi=\pi$.}
\end{figure}

Let $P_\phi(m)$ be the probability of finding the particle on an edge connected to the special vertex after one repetition of the experiment assuming that $m$ steps of the walk are taken before the measurement, and the phase shift is $\phi$. Then the average number of steps $\bar n_{\phi,m}$ to be taken when measuring after $m$ steps on a graph with phase shift $\phi$ is given by
\[
\bar n_{\phi,m}=\sum_{k=1}^\infty P_{\phi,m}(k) km,
\]
where
\[
P_{\phi,m}(k)=[1-P_\phi(m)]^{k-1}P_\phi(m)
\]
is the probability of finding the particle on an edge connected to the special vertex after $k$ repetitions of the experiment (and no sooner). After a short evaluation we find
\[
 \bar n_{\phi,m}=\frac{m}{P_\phi(m)}.
\]
The optimal number of steps, $\opt$, in each experiment is given by the value of $m$ that minimizes $\bar n_{\phi,m}$.  We will denote the average number of steps to be taken in the optimal case as $\bar n(\phi)=\bar n_{\phi,\opt}$.  This is plotted in Fig.~\ref{fig:optimality}.

For $\phi=\pi$, the quantum approach is faster than the classical; it has a quadratic speedup, even in the nonoptimal search via the maximum probability method (see Fig.~\ref{fig:optimality}). However, for the case of $\phi\sim0$ we see that the optimal number of steps before measuring is $0$---we do not evolve the system, we just measure the initial state---and the average number of steps to find the special vertex reaches a value close to $N$. This is the same situation we would have if we performed a classical blind search in which we randomly pick vertices at each step without remembering the past choices.

\subsection{Bipartite graph}
\label{sec:bipartite}

\begin{figure}
\includegraphics[scale=0.9]{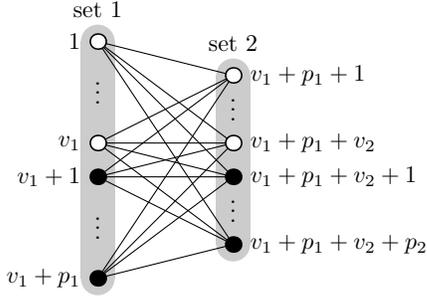}
\caption{\label{fig:bipG}General bipartite graph having $v_1$ special (black circles) and $p_1$ normal (white circles) vertices in the first set and $v_2$ special and $p_2$ normal vertices in the second set. This problem can be reduced to an eight-dimensional one with a two-dimensional decoupled subspace. The remaining six dimensions can be reduced to three by performing two steps at a time due to the oscillatory behavior of the walk in bipartite graphs.}
\end{figure}

Let us now consider a bipartite graph consisting of two sets of vertices in which each vertex in each set is connected to every vertex in the other set, but there is no connection between vertices in the same set. Let us further suppose there are $N_{1}$ vertices in the set $1$ and $N_{2}$ in the set $2$. Of all vertices in the set $1$, $v_{1}$ are special vertices and $p_{1}=N_{1}-v_{1}$ are normal vertices, and in the set $2$, $v_{2}$ are special vertices and $p_{2}=N_{2}-v_{2}$ are normal vertices---see Fig.~\ref{fig:bipG}.  The action of the unitary operator, $\op U$ acting on a state entering a normal vertex in the set $1$ is given by Eq. (\ref{eq:unitary}) with\footnote{The subscripts of  the  transmission and reflection coefficients indicate to which set the corresponding vertex does belong.}
\[
t_{1}=\frac{2}{N_{2}}, \hspace{5mm} r_{1}=\frac{N_{2}-2}{N_{2}}.
\]
If it acts on a state entering a normal vertex in the set $2$ its action is again given by Eq. (\ref{eq:unitary}) but with
\[
t_{2}=\frac{2}{N_{1}}, \hspace{5mm} r_{2}=\frac{N_{1}-2}{N_{1}}.
\]
The action of $\op U$ on a state entering a special vertex is given by Eqs.~(\ref{eq:unitary}) and (\ref{eq:RC}).  We shall number the vertices in the set $1$ as $1$ through $N_{1}$ and those in the set $2$ as $N_{1}+1$ through $N_{1}+N_{2}$.  In analyzing the walk, we can, without loss of generality, assume that vertices $1$ through $v_{1}$ in the set $1$ are special, and vertices $N_{1}+1$ through $N_{1}+v_{2}$ are special as depicted on Fig.~\ref{fig:bipG}.

There are now eight invariant subspaces, two of which, consisting of edges that connect the sets of special vertices, are decoupled from the rest. So our problem is essentially six-dimensional.  We can see this by first defining the vectors
\begin{eqnarray*}
 |w_{01}\rangle & = & \frac{1}{\sqrt{v_{1}v_{2}}}\sum_{j=1}^{v_{1}}; 
\sum_{k=N_{1}+1}^{N_{1}+v_{2}} |k,j\rangle, \nonumber  \\
|w_{02}\rangle & = & \frac{1}{\sqrt{v_{1}v_{2}}}\sum_{j=1}^{v_{1}}
\sum_{k=N_{1}+1}^{N_{1}+v_{2}} |j,k\rangle .
\end{eqnarray*}
The vector $|w_{01}\rangle$ is a superposition of states leaving special vertices in the set $2$ and entering special vertices in the set $1$, and $|w_{02}\rangle$ is a superposition of edge states leaving special vertices in the set $1$ and entering special vertices in the set $2$.  We have that $\op U |w_{01}\rangle = \exp (i\phi ) |w_{02}\rangle$ and $\op U |w_{02}\rangle = \exp (i\phi ) |w_{01}\rangle$, so that these vectors decouple from the rest of the problem.

Now let us define
\begin{eqnarray*}
|w_{11}\rangle & = & \frac{1}{\sqrt{v_{1}p_{2}}} \sum_{j=1}^{v_{1}}
\sum_{k=N_{1}+v_{2}+1}^{N_{1}+N_{2}} |k,j\rangle  \nonumber , \\
|w_{12}\rangle & = & \frac{1}{\sqrt{v_{2}p_{1}}} \sum_{j=v_{1}+1}^{N_{1}}
\sum_{k=N_{1}+1}^{N_{1}+v_{2}} |k,j\rangle  \nonumber  , \\
|w_{13}\rangle &= & \frac{1}{\sqrt{p_{1}p_{2}}}  \sum_{j=v_{1}+1}^{N_{1}}
\sum_{k=N_{1}+v_{2}+1}^{N_{1}+N_{2}} |k,j\rangle .
\end{eqnarray*}
These vectors consist of different sets of edge states entering the set $1$.  Let us denote their span by $\subspace_{1}$.  Let us define the vectors
\begin{eqnarray*}
|w_{21}\rangle & = & \frac{1}{\sqrt{v_{1}p_{2}}} \sum_{j=1}^{v_{1}}
\sum_{k=N_{1}+v_{2}+1}^{N_{1}+N_{2}} |j,k\rangle  \nonumber ,  \\
|w_{22}\rangle & = & \frac{1}{\sqrt{v_{2}p_{1}}} \sum_{j=v_{1}+1}^{N_{1}}
\sum_{k=N_{1}+1}^{N_{1}+v_{2}}  |j,k\rangle \nonumber , \\
|w_{23}\rangle & = & \frac{1}{\sqrt{p_{1}p_{2}}} \sum_{j=v_{1}+1}^{N_{1}}
\sum_{k=N_{1}+v_{2}+1}^{N_{1}+N_{2}} |j,k\rangle  .
\end{eqnarray*}
These vectors consist of different sets of edge states entering the set $2$.  We shall denote their span by $\subspace_{2}$.

Finally, let us consider the action of the unitary operator that advances the walk one step on these states. We find that
\begin{eqnarray*}
\op U |w_{11}\rangle & = & e^{i\phi} |w_{21}\rangle , \\
\op U |w_{12}\rangle & = & [t_{1}(v_{2}-1)-r_{1}] |w_{22}\rangle + t_{1}\sqrt{v_{2}p_{2}}
 |w_{23}\rangle   \nonumber  , \\
\op U |w_{13}\rangle & = & [t_{1}(p_{2}-1)-r_{1}] |w_{23}\rangle + t_{1}\sqrt{v_{2}p_{2}}
|w_{22}\rangle, \nonumber
\end{eqnarray*}
and
\begin{eqnarray*}
\op U |w_{21}\rangle & = & [t_{2}(v_{1}-1) - r_{2}] |w_{11}\rangle + t_{2}\sqrt{v_{1}p_{1}}
|w_{13}\rangle  \nonumber  , \\
\op U |w_{22}\rangle & = & e^{i\phi} |w_{12}\rangle , \\
\op U |w_{23}\rangle & = & [t_{2}(p_{2}-1) - r_{2}] |w_{13}\rangle + t_{2}\sqrt{v_{1}p_{1}}
|w_{11}\rangle .\nonumber
\end{eqnarray*}
From these equations, we see that $\op U$ maps $\subspace_{1}$ into $\subspace_{2}$ and vice versa.  This also implies that $\op U^2$ maps $\subspace_{1}$ into itself and $\subspace_{2}$ into itself.  Therefore, if we consider a walk with an even number of steps, our six-dimensional problem turns into two three-dimensional ones.  This is what we shall do.

We shall look in detail at what happens in $\subspace_{2}$; the case of $\subspace_{1}$ is similar.  We shall only consider the case $\phi = \pi$.  First, let us define the quantities
\begin{eqnarray*}
q_{1} = -r_{1}+t_{1}(v_{2}-1), & s_{1} = t_{1}\sqrt{v_{2}p_{2}},   \nonumber   \\
q_{2} = -r_{2}+t_{2}(v_{1}-1), & s_{2} = t_{2}\sqrt{v_{1}p_{1}},
\end{eqnarray*}
and note that $q_{j}^{2}+ s_{j}^{2} = 1$ for $j=1,2$. If we denote $\op U^2$ restricted to $\subspace_{2}$ by $\op M$, then the matrix for $\op M$ in the basis $|w_{21}\rangle$, $|w_{22}\rangle$, and $|w_{23}\rangle$ reads
\[
\op M = \begin{pmatrix} -q_{2} & 0 & -s_{2} \\ s_{1}s_{2} & -q_{1} & -q_{2}s_{1} \\
-q_{1}s_{2} & -s_{1} & q_{1}q_{2} \end{pmatrix}.
\]
In finding the eigenvalues and eigenvectors of this matrix, we assume that the number of special vertices is small, that is, $(v_{1}/N_{1}) \ll 1$ and $(v_{2}/N_{2})\ll 1$, and we further neglect higher-order terms. In addition, we set $x_{1}=2v_{1}/N_{1}$ and $x_{2}=2v_{2}/N_{2}$.

For the initial state of our walk, we will choose the state that is an equal superposition of all edge states entering the set $2$. The behavior of the walk is similar for any initial state that is a linear combination of this state and the state that is an equal superposition of all edge states entering the set $1$.  Note that the state that is an equal superposition of all edge states in the graph is of this form.  For simplicity we choose to start the walk in the equal superposition of all edge states entering the set $2$. In this case the entire walk, to very good approximation, takes place in the subspace $\subspace_{2}$. Our initial state can be expressed as
\begin{eqnarray*}
|\psinit\rangle & = & \frac{1}{\sqrt{N_{1}N_{2}}}(\sqrt{v_{1}v_{2}}|w_{02}\rangle + \sqrt{v_{1}p_{2}}
|w_{21}\rangle \nonumber \\
 & & + \sqrt{v_{2}p_{1}} |w_{22}\rangle + \sqrt{p_{1}p_{2}} |w_{23}\rangle ) .
\end{eqnarray*}
This vector is not entirely in $\subspace_{2}$ due to the $\ke{w_{02}}$ component. But the component that is not in $\subspace_{2}$ is small, and it stays small throughout the evolution.  Neglecting this small component, we find that
\[
\op U^{2n} |\psinit\rangle \simeq \begin{pmatrix} -\sqrt{x_{1}/(x_{1}+x_{2})} \sin (2n\theta ) \\
\sqrt{x_{2}/(x_{1}+x_{2})} \sin (2n\theta ) \\ -\cos (2n\theta ) \end{pmatrix}.
\]
From this equation, we see that, when
\begin{equation}
\label{bipartitestepnum}
2n = \frac{\pi}{2\theta} = \frac{\pi}{2\sqrt{2(x_{1}+x_{2})}} ,
\end{equation}
we are with certainty on an edge connected to a special vertex.  If after this many steps we measure the particle to determine which edge it is on, with probability $x_{1}/(x_{1}+x_{2})$ we find it on an edge connected to a special vertex in the set $1$, and with probability $x_{2}/(x_{1}+x_{2})$ we find it on an edge connected to a special vertex in the set $2$.

In order to get a better understanding for this solution, let us consider the case $v_{1}=v_{2}=1$.  In this case
\[
\theta = 2\left( \frac{1}{N_{1}} + \frac{1}{N_{2}}\right)^{1/2} .
\]
When $2n = \pi /2\theta$, the probability of finding the particle on an edge connected to the special vertex in the set 1 is $N_{2}/(N_{1}+N_{2})$ and the probability of being on an edge connected to the special vertex in the set $2$ is $N_{1}/(N_{1}+N_{2})$.  Now let us suppose that $N_{2} \gg N_{1}$ and ask the question: How many steps would it take to find the special vertices in each of the sets? We find that the number of steps in the walk is $\pi /2\theta \sim \sqrt{N_{1}}$.  The number of times the walk would have to be repeated in order to find the special vertex in the set $1$ is $\order (1)$, while the number of times to find the special vertex in the set $2$ is $\order (N_{2}/N_{1})$. Therefore, the total number of steps to find the special vertex in the set $1$ is $\order (\sqrt{N_{1}})$, and the total number of steps to find the special vertex in the set $2$ is $\order (N_{2}/\sqrt{N_{1}})$.  A classical search would require $\order (N_{1})$ steps to find the special vertex in the set $1$ and $\order (N_{2})$ steps to find the special vertex in the set $2$.

Another interesting case is when all of the special vertices are only in one set. For example, if they are all in the set $1$, then $x_{2}=0$, and we see from Eq.~(\ref{bipartitestepnum}) that the number of steps that concentrates the probability on edges connected to special vertices is $\order (\sqrt{N_{1}/v_{1}})$, which does not depend on the number of vertices in the set $2$.  Therefore, if the special vertices are only in one set, the size of the other set does not affect the number of steps necessary to find a special vertex.

\subsection{$M$-partite complete graph}
\label{sec:mpartite}

\begin{figure}
\includegraphics[scale=0.7]{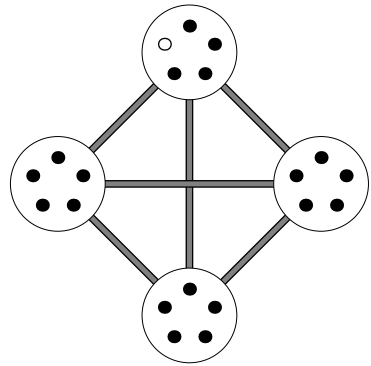}
\caption{\label{fig:mpartite} $M$-partite complete graph consists of $M$ sets of vertices where each set contains $N$ vertices. For every pair of vertices not belonging to the same set there exists an edge, while there is no edge connecting any two vertices within the same set. In one of the sets we consider one vertex to be special.}
\end{figure}

As a generalization of previous cases we may consider an $M$-partite complete graph consisting of $M$ sets of $N$ vertices. Vertices within sets are not connected while for any pair of vertices not belonging to the same set there exists an edge connecting them (see Fig.~\ref{fig:mpartite}). We see that by taking $M=2$ we obtain the bipartite graph discussed in Sec.~\ref{sec:bipartite}. However, if we suppose that there is only one vertex in each set, we get a complete graph, which was discussed in Sec.~\ref{sec:cg}. In both of these cases, a quadratic speedup of classical searches has been demonstrated. For the graph we are presently considering, $M$ is its chromatic number. The chromatic number of a graph is the number of colors one would need to color its vertices so that no two vertices connected by an edge have the same color.

To simplify the analysis, let us suppose, that there is only one special vertex and both $M$ and $N$ are large (see Fig.~\ref{fig:mpartite}). We will label the vertices by $(m,n)$, where $m=1,2,\ldots, M$ is the number labeling a group (set), and $n$ is the number labeling vertices within the group. We shall further assume that the special vertex is $(1,1)$. It turns out that this problem is five-dimensional, where the subspace $\subspace$ in which the walk is happening is defined by five vectors according to Eq.~(\ref{eq:eigenone}): $\ke{w_1}$ consisting of edges entering the special vertex, $\ke{w_2}$ consisting of edges leaving the special vertex, $\ke{w_3}$ and $\ke{w_4}$ consisting of the edges leaving and entering normal vertices in the set $1$, respectively, and $\ke{w_5}$ consisting of all of the edges connecting vertices in sets $2$ through $M$. Note that in the case $M=2$ we do not have the vector $\ke{w_5}$.

The unitary operator $\op U$ advancing the walk by one step is again defined by Eqs.~(\ref{eq:unitary}) and (\ref{eq:RC}) for the special vertex and by Eqs.~(\ref{eq:unitary}) and (\ref{eq:GC}) for normal vertices with
\[
t=\frac{2}{N(M-1)},\qquad r=1-t.
\]

By analyzing the action of $\op U$ on vectors $\ke{w_k}$ for $k=1,2,\ldots ,5$ (see Appendix B), neglecting higher-order terms, and, finally, employing Eq.~(\ref{eq:evolution}) we find that
\[
\op U^n\ke\psinit\simeq\begin{pmatrix}
\sin\phi n\\
-\sin\phi n\\
0\\
0\\
\sqrt{2}\cos\phi n
\end{pmatrix},
\]
where $\cos\phi\simeq 1-1/MN$. We observe, that if $\phi n=\frac{\pi}{2}$, only the first two components are nonzero, leading to the probability of finding the sought element being close to 1. In this case $n\sim\sqrt{MN}$, which is again quadratic speedup over classical search, in which $\order(MN)$ steps would be needed.

\subsection{Comparison with other works}

There are several papers (to the authors' knowledge) that have also studied searches on graphs. We would like to compare the results in these papers with our results.

The problem discussed here is, in fact, similar to the one in Ref.~\cite{ShKeWh03}, where a search on a hypercube was studied. There an algorithm for finding the (only one) special vertex with a probability of approximately $1/2$ was presented, which is in contrast with our finding of a probability close to unity. In our case the probability is (almost) equally split between two possible sets of edge states, those \emph{leaving} the special vertex and those \emph{entering} the special vertex. In the case of Ref.~\cite{ShKeWh03}, this corresponds to either finding the particle on the special vertex (with an arbitrary coin state) or finding it on one of the neighboring vertices with the coin pointing to the special one. The authors of Ref.~\cite{ShKeWh03} discard the latter type of states as an unsuccessful search. Similar findings are given in Ref.~\cite{PoGaKiJe08}.

In Ref.~\cite{AmKeRi05} a quantum walk that performs an exact Grover search is discussed. The graph the quantum walk is performed on is a complete graph that has loops added to each vertex. Coins (equivalents of our local unitary evolutions) for normal vertices are Grover coins. However, coins for the special vertices are ``minus'' Grover coins. This leads to almost the same evolution as in the case of a Grover search; the only difference is that one step of the Grover algorithm corresponds to two steps of the quantum walk.

Finally, let us summarize the situation as it presently stands. Quantum-walk searches have two figures of merit, the number of steps necessary to find a special vertex and the probability of finding it after a specific number of steps have been made.  On a complete graph without loops [the  result obtained in this paper in Eq.~(\ref{eq:result})] one needs $\sqrt{2}$-times as many steps as in the Grover search for the corresponding problem (a search on $N$ objects), and after this many number of steps the probability of finding the special vertex is equal to unity. On a complete graph with loops \cite{AmKeRi05}, twice as many steps as in the corresponding Grover search are required, and the probability of finding the special vertex is again equal to unity. A rigorous comparison of these properties on a hypercube can be found in Ref.~\cite{PoGaKiJe08}, where adding loops to the graph incurs the cost of increasing the number of steps in the walk.

\section{Conclusion}
\label{sec:conclusion}

We have considered scattering-quantum-walk searches on several examples of highly symmetric (complete, bipartite, and $M$-partite) graphs where some of the vertices are special.  We have shown how an oracle in a quantum circuit can be used to mark these special vertices. The symmetry of these graphs leads to a significant reduction in the dimensionality of the problem. For all of the types of graphs we considered, we found a quadratic speedup over a classical search. These results were obtained by taking the phase shift of special vertices to be $\pi$. Taking $\phi=0$ results in a trivial evolution, with constant probabilities of finding the particle in any edge state. In this case quantum walks reduce to the classical blind search on edges.  That is, the quantum search is reduced to its classical counterpart.

We have also studied the change in behavior when changing the phase shift $\phi$ for a complete graph with one special vertex. While for $\phi=\pi$ we have a nontrivial behavior suitable for searches in these graphs, for $\phi=0$ we get only a static walk. Cases between these values were explored numerically. As a measure for suitable comparison between different choices of $\phi$, we chose the average number of steps that need to be taken to successfully find the special vertex. We note that the quantum algorithm is always at least as fast as its classical counterpart.

Finally, we have made comparisons of our results with other works, putting this work into perspective. It is hoped that the approach used here makes the evaluation of these cases easier and more simple to interpret.

\section*{Acknowledgments}
Our work has been supported by projects the QAP Project No. 2004-IST-FETPI-15848, the HIP Project No. FP7-ICT-2007-C-221889 and the Project APVV QIAM.

\appendix
\section{Classical searches}
\label{sec:probabilities}

When employing quantum oracles in searches for an element in an unstructured database, the classical search may be viewed as a quantum query not capable of using superpositions. In this way one may consider two types of classical searches---blind searches and searches with a memory. We present both types of search below.

\subsection{Blind searches}
\label{sec:bs}

Classical \emph{blind searches} are searches where previously chosen elements are not marked, so that they may be picked up again. This means that  the probability $P$ of choosing a special element remains the same after every unsuccessful step of the search. The probability of finding a special element after $k$ steps, hence, is
\[
P_k=(1-P)^{k-1}P,
\]
which is the probability of not finding the element in $k-1$ steps and finding it at the $k$-th step. The average number of steps we need to take in order to find a marked element is expressed as usual:
\[
\bar n=\sum_{k=1}^\infty P_kk.
\]
In the case when the oracle function marks $v$ vertices out of $N$, we have $P=v/N$ so that
\[
\bar n=\frac{1}{P}=\frac{N}{v}.
\]

\subsection{Classical searches with memory}
\label{sec:ms}

Let us suppose that we mark previously chosen elements, and do not choose them again in subsequent steps during the search. The probabilities $P_k$ of finding a marked element now change for each new element chosen:
\begin{eqnarray*}
P_k & = & \biggl[\frac{N-v}{N}\cdot\frac{N-v-1}{N-1}\cdots \\
& & \cdots\frac{N-v-k+2}{N-k+2}\biggr]\cdot\frac{v}{N-k+1} \\
& = & \frac{(N-v)!v}{N!}\cdot\frac{(N-k)!}{(N-v-k+1)!}.
\end{eqnarray*}
This is a product of $(k-1)$ probabilities of not finding a special vertex in successive searches, where after each unsuccessful search we remove the selected normal vertex from the search. In this case, the average number of steps taken in order to find a marked vertex is
\[
\bar n=\sum_{k=1}^{N-v+1}P_k  k .
\]
After some evaluation this yields
\[
\bar n=\frac{N+1}{v+1}\sim\frac{N}{v}.
\]
In both cases we may observe that the efficiency remains the same, of the order $N/v$.

\section{$M$-partite graph}
The explicit expressions for the vectors $|w_{1}\rangle$, \ldots, $|w_{5}\rangle$ read 
\begin{eqnarray*}
|w_{1}\rangle & = & \frac{1}{[N(M-1)]^{1/2}} \sum_{m=2}^{M} \sum_{n=1}^{N} |(m,n),(1,1)\rangle
\nonumber , \\
|w_{2}\rangle & = &  \frac{1}{[N(M-1)]^{1/2}} \sum_{m=2}^{M} \sum_{n=1}^{N} |(1,1),(m,n)\rangle
\nonumber , \\
|w_{3}\rangle & = & \frac{1}{[N(N-1)(M-1)]^{1/2}}\\
& & \qquad\times \sum_{m=2}^{M} \sum_{n_{1}=2}^{N}
\sum_{n_{2}=1}^{N} |(1,n_{1}),(m,n_{2})\rangle  \nonumber , \\
|w_{4}\rangle & = & \frac{1}{[N(N-1)(M-1)]^{1/2}} \\
& & \qquad\times \sum_{m=2}^{M} \sum_{n_{1}=2}^{N}
\sum_{n_{2}=1}^{N} |(m,n_{2}),(1,n_{1})\rangle  \nonumber , \\
|w_{5}\rangle & = & \frac{1}{[N(M-1)(M-2)]^{1/2}} \sum_{m_{1}=2}^{M}
\sum_{\substack{m_{2}=2\\m_{2}\neq m_{1}}}^{M}  \nonumber  \\
& & \qquad\times \sum_{n_{1}=1}^{N} \sum_{n_{2}=1}^{N} |(m_{1},n_{1}),(m_{2},n_{2})\rangle  .
\end{eqnarray*}
The actions of $\op U$ on these vectors are
\begin{eqnarray*}
U|w_{1}\rangle & = & e^{i\phi}|w_{2}\rangle  \nonumber ,  \\
U|w_{2}\rangle & = & -r|w_{1}\rangle + t\sqrt{N-1}|w_{4}\rangle + t \sqrt{N(M-2)} |w_{5}\rangle
\nonumber , \\
U|w_{3}\rangle & = & t\sqrt{N-1} |w_{1}\rangle - \frac{N(M-3)+2}{N(M-1)} |w_{4}\rangle \\
& & + t [N(N-1)(M-2)]^{1/2} |w_{5}\rangle \nonumber , \\
U|w_{4}\rangle & = & |w_{3}\rangle \nonumber  , \\
U|w_{5}\rangle & = & t \sqrt{N(M-2)} |w_{1}\rangle \\
& & + t [N(N-1)(M-2)]^{1/2} |w_{4}\rangle
+\frac{M-3}{M-1} |w_{5}\rangle .
\end{eqnarray*}
These equations give us a $5\times 5$ matrix, whose eigenvalues in the large-$M$ and large-$N$ limit are approximately $-1$, $\pm i -(1/M)$, and $1\pm i\sqrt{t}$, respectively. Only the eigenvectors corresponding to the eigenvalues $1\pm i\sqrt{t}$ have a significant overlap with the initial state, so it is only these eigenvalues that determine the behavior of the walk.

\end{document}